# CFD Simulation of the NREL Phase VI Rotor

Y. Song* and J. B. Perot[#]

*Theoretical & Computational Fluid Dynamics Laboratory, Department of Mechanical & Industrial Engineering, University of Massachusetts Amherst, Amherst, MA 01003, USA

[#]Theoretical & Computational Fluid Dynamics Laboratory, Department of Mechanical & Industrial Engineering, University of Massachusetts Amherst, Amherst, MA 01003, USA

[#]E-Mail: perot@acad.umass.edu (Corresponding Author)

**ABSTRACT**: The simulation of the turbulent and potentially separating flow around a rotating, twisted, and tapered airfoil is a challenging task for CFD simulations. This paper describes CFD simulations of the NREL Phase VI turbine that was experimentally characterized in the 24.4m × 36.6m NREL/NASA Ames wind tunnel (Hand et al., 2001). All computations in this article are performed on the experimental base configuration of $0^o$ yaw angle, $3^o$ tip pitch angle, and a rotation rate of 72 rpm. The significance of specific mesh resolution regions to the accuracy of the CFD prediction is discussed. The ability of CFD to capture bulk quantities, such as the shaft torque, and the detailed flow characteristics, such as the surface pressure distributions, are explored for different inlet wind speeds. Finally, the significant three-dimensionality of the boundary layer flow is demonstrated.

## 1. INTRODUCTION

Wind turbine blades often operate outside their design conditions. When this happens, the flow field and resulting forces and torques on the blade can become quite complex. The turbulent boundary layer becomes three-dimensional and often separates entirely from the airfoil. A detailed understanding of the flow conditions around the blades could substantially reduce the fatigue on the blades and the downstream transmission system, and enable control mechanisms to be implemented. CFD has the potential to produce this level of flow prediction.

The simulations in this work use the open-source CFD code, OpenFOAM (OpenFOAM, 2011). They are therefore reasonably indicative of what could be expected from most publically available CFD codes. The simulations in this work use the Spalart-Allmaras turbulence model (Spalart and Allmaras, 1994). This model was developed at Boeing specifically for airfoil predictions. In an effort to predict boundary layer separation correctly, wall functions are not used. The partial differential equations are integrated completely to the wall. This means that the first grid point away from the airfoil always resides within the laminar sub-layer of the turbulent boundary layer. At the high Reynolds numbers ($10^5$) encountered on turbine blades, the boundary layers are very thin, and this is a numerically challenging problem.

The CFD predictions are validated against the experiments presented in the NREL report by Robinson et al. (1999). The NREL Phase VI test is a full scale Unsteady Aerodynamic experiment (UAE) on the double-bladed 10.058 m diameter NREL Phase VI Rotor based on S809 airfoil and performed in the 24.4 m × 36.6 m NASA-Ames wind tunnel (Hand et al., 2001). This experiment has been used by a wide variety of prior studies. Tangler (2002) tested multiple versions of a Blade Element Method code. Laino et al. (2002) performed a 2D simulation of the S809 airfoil using the AERODYN code and matches those results with the NREL data. Sorensen et al. (2002) applied an incompressible RANS code to predict several cases form the NREL and NASA wind tunnel tests. Duque et al. (2003) gave a comprehensive investigation of a RANS computation using the CAMRAD II and OVERFLOW-D2 codes performed on the double-blade NREL Phase VI rotor. Xu and Sankar (2000) performed a RANS computation using a 3D viscous flow model. Gonzalez and Munduate (2008) analyzed the aerodynamic properties of the blades, such as attached flow, separated flow, and stall, of parked and rotating configurations of the NREL Phase VI wind turbine by testing a 2D section of the blades. Similar results for the same configuration were also presented by Schmitz and Chattot (2006). This work differs from those prior studies in its choice of turbulence model (Spalart-Almaras), and the decision not to use wall-function boundary conditions which algebraically model the boundary layer profile.

## 2. CFD MODELING

### 2.1 Method

In the present work, an implicit transient solver is applied to predict the aerodynamics of the UAE Phase VI rotor. The double-bladed 10.058 m diameter Phase VI rotor geometry is based on a twisted and tapered S809 airfoil. The specifications can be found in Hand et al., (2001). The turbine is set with $0^o$ yaw angle and $3^o$ tip pitch angle and with a rotation rate of 72 rpm. The downwind tower and the nacelle are not included in these simulations since their aerodynamic effects on the blades can largely be neglected. The simulated tunnel size is the same as the 24.4 m × 36.6 m NREL/NASA Ames wind tunnel.

The software that is used in this research is OpenFOAM-1.6-ext (OpenFOAM, 2011). The algorithm used to solve the Reynold's Averaged Navier-Stokes equations is PIMPLE, an incompressible transient turbulent flow solver, which combines the PISO and SIMPLE algorithms for computing the pressure. The PIMPLE algorithm is compiled in the OpenFOAM solver, pimpleDyMFoam, and is used in all the computations presented herein. PISO stands for Pressure Implicit with Splitting the Operators algorithm while SIMPLE represents Semi-Implicit Method for Pressure-Linked Equation algorithm. For a detailed explanation of the PISO and SIMPLE algorithms see Jasak (1996).

In each simulation, air flows in the inlet of the tunnel, which is the inertial frame and some of the air then goes through a refined mesh that resides in a cylinder. The blade is contained in this cylinder and the cylinder mesh rotates with respect to the outer wind tunnel mesh. The interface between the two meshes is handled by a special boundary condition called the Generalized Grid Interface (GGI) (Schmitt, 2009). The GGI interface uses a special interpolation algorithm that allows for general grid movement. This arrangement avoids the complicated work of a topologically deforming mesh. In

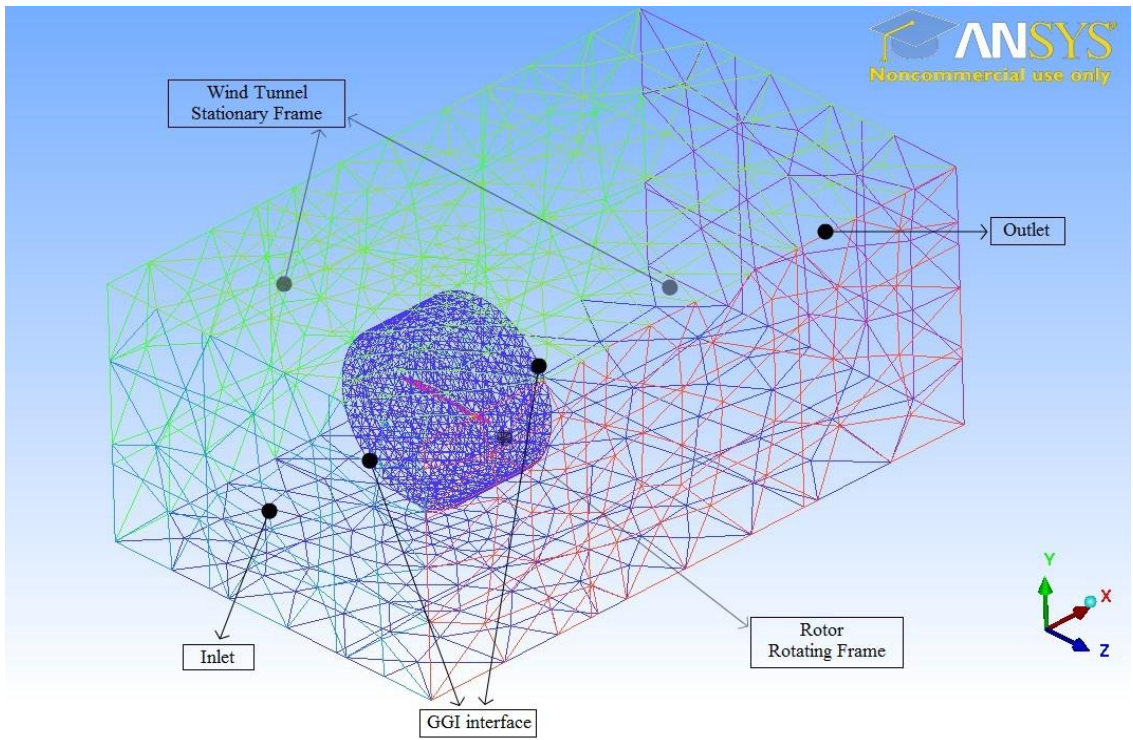

**Figure 1.** Surface mesh from an unstructured mesh over the NREL Phase VI blade geometry. Blade is the small strip object. Rotating mesh is the cylinder. Outer mesh is the wind tunnel. This is a coarse version of the mesh, not the final one.

Figure 1 the cylinder mesh rotates inside the larger cuboid mesh. And the blade geometry is embedded in the cylinder mesh.

All the computations are performed on the Theoretical and Computational Fluid Dynamics Laboratory constructed supercomputing cluster, Cyclops, which has 608 processing cores. A typical simulation, which uses an unstructured mesh that contains 10 million cells, takes 48 hours per revolution of the blades using 128 cores on Cyclops. This is about 0.2 milliseconds per gridcell per timestep.

**2.2 Mesh Refinement**

Unstructured meshes are applied throughout this project so that the CFD simulation can be performed on the wind turbine blade which has a complex geometry. Figure 2 shows the surface mesh at the trailing edge of the NREL Phase VI turbine blade. The mesh around the sharp trailing edge of the blade consists of cells with severely non-orthogonal faces. This can make the CFD solution unstable or inaccurate. In order to get rid of those highly non-orthogonal cell faces, slight modification of the turbine geometry is made by making the sharp trailing edge blunt. Figure 3 shows a zoomed in picture of a section of the mesh around the modified trailing edge. This modification reduces the chord length by roughly 2%.

In order to avoid using wall functions the mesh is highly refined in the wall normal direction in a thin layer next to the airfoil. In order to integrate the PDE without wall function boundary conditions the first grid point away from the wall needs to reside at a $y^+$ less than or equal to 5. High aspect ratio tetrahedra have very non-orthogonal faces, so next to the airfoil the mesh consists of very flat prisms.

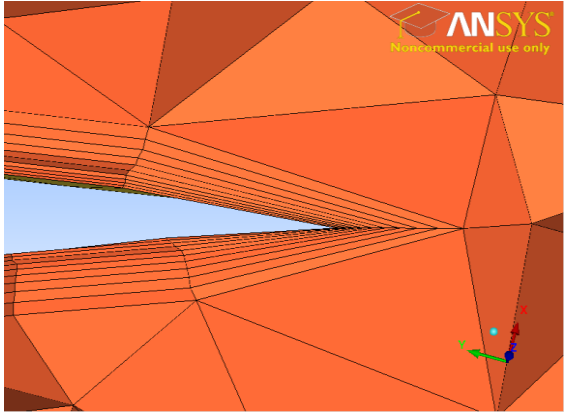 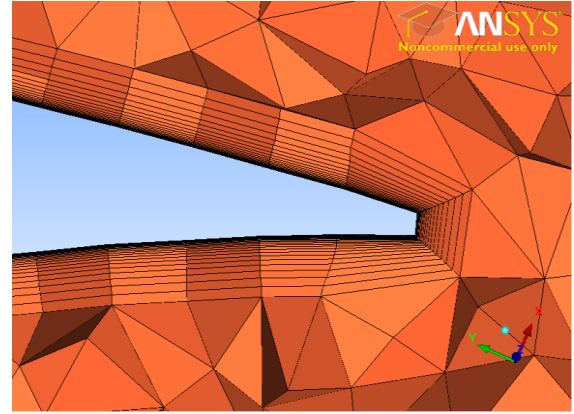

**Figure 2.** A sliced section of the mesh around the sharp trailing edge (before modified)

**Figure 3.** A section of the mesh around the sharp trailing edge (after modification)

The prisms are aligned normal to the blade surface. In figures 2 and 3, the rectangles near the airfoil are actually a slice through these prisms. To integrate the equations up to the wall, the first prism layer next to the wall needs to be $5\times10^{-5}$ meters high. The thickness of each prism grows by 15% as the prisms move away from the wall. In many locations the prisms next to the airfoil are very thin and have an aspect ratio (height to width ratio) of over 400.

Normal resolution is required on the entire blade, but chordwise and spanwise resolutions are also required at the blade leading edge. At this location there are large pressure tangential pressure gradients that must be resolved for an accurate computation. Meshes with different leading edge resolutions were investigated. Figures 4, 5 and 6 show sliced sections of the mesh at 80% span of the blade, which contain a total of 4 million, 6.3 million and

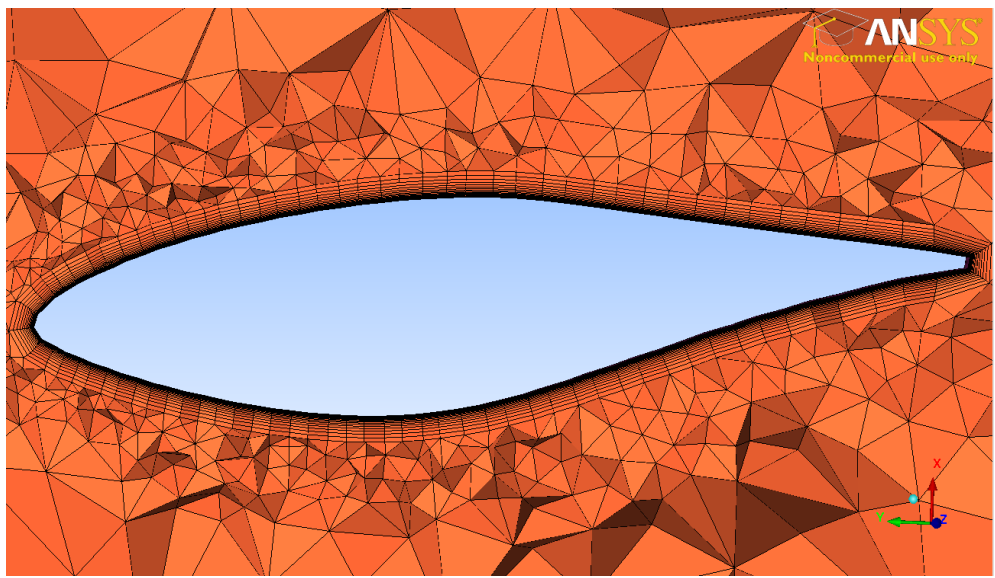

**Figure 4.** A sliced section of the mesh at 80% span of the NREL Phase VI blade with 4 million total mesh cells and a coarse resolution of the leading edge.

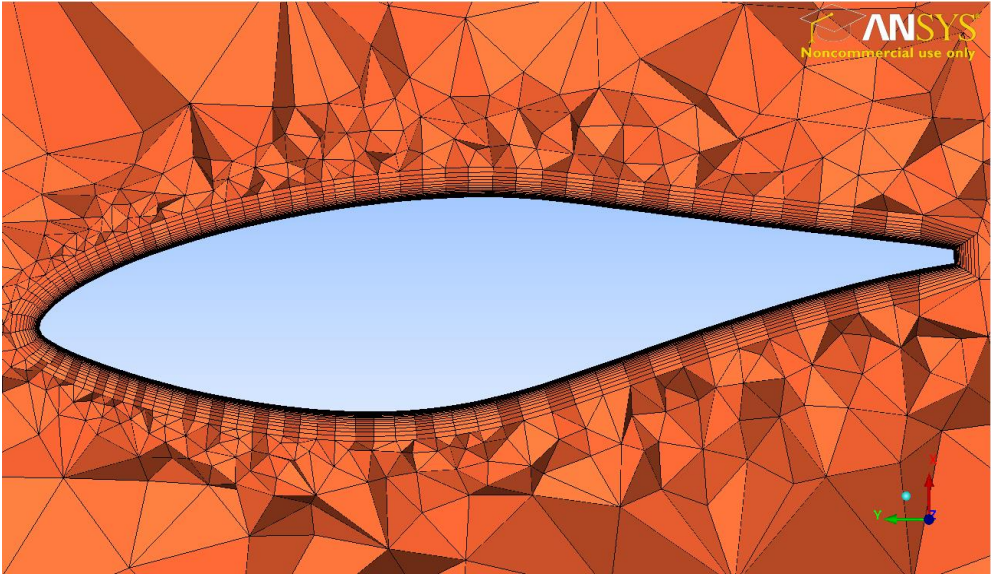

**Figure 5**. A sliced section of the mesh at 80% span of the NREL Phase VI blade with 6.3 million cells total and almost twice the mesh resolution on the leading edge.

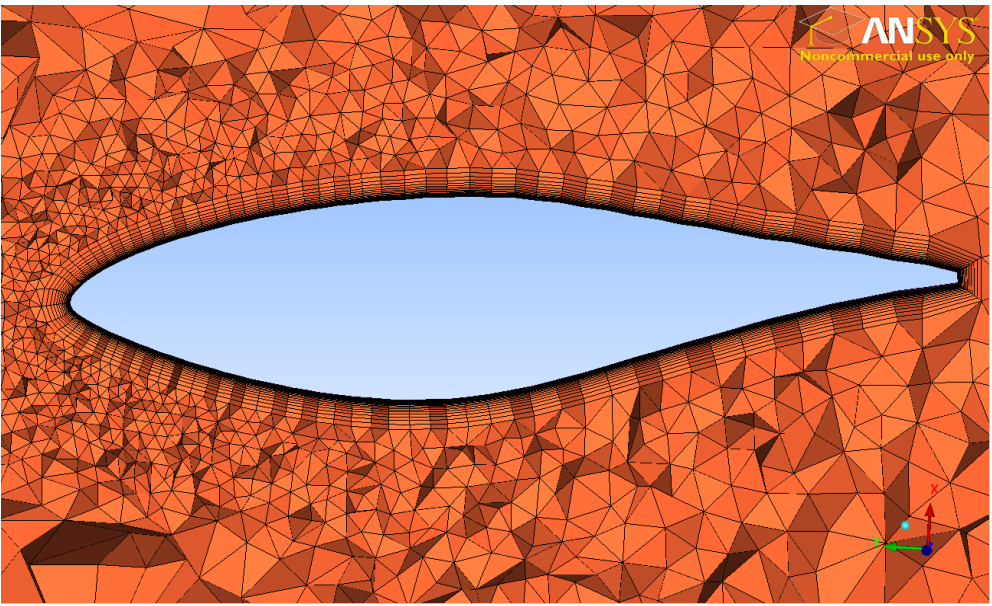

**Figure 6.** A sliced section of the mesh at 80% span of the NREL Phase VI blade with 10 million cells total and sufficient leading edge resolution.

10 million mesh cells respectively. Figure 7 shows the pressure distributions computed for these 3 different meshes. These results are for the inlet wind speed of 10m/s at 80% span. Only the largest mesh is capable of predicting the pressure spike at the leading edge (and the resulting blade torque) reasonably accurately.

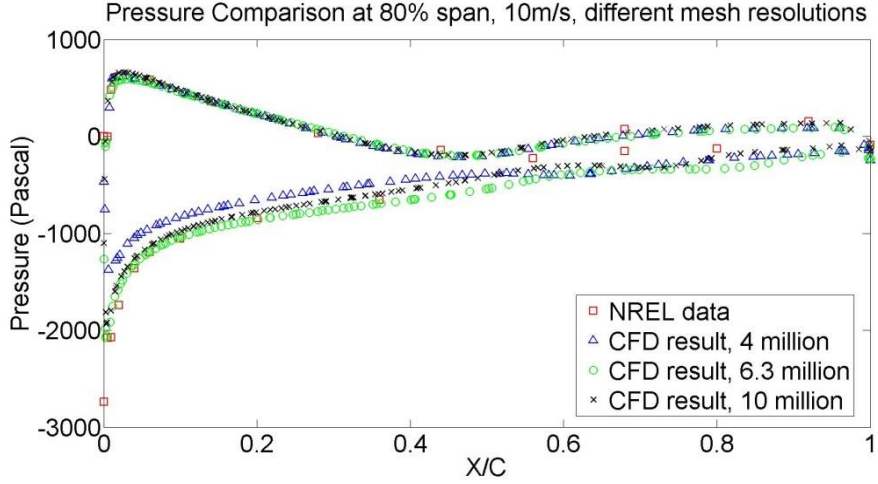

**Figure 7.** Comparison of pressure distributions with different mesh resolutions at 80% span for the 10m/s case

### 2.3 Boundary conditions

For all the simulation presented in this work, the pressure is enforced as zero gradient at the inlet of the tunnel and zero value at the outlet, while the velocity is fixed at the inlet and has a zero gradient boundary condition at outlet. Slip conditions are used at the four side walls of the wind tunnel (so the thin boundary layers on the tunnel walls are not captured), and the no-slip boundary condition is applied on the blade surface. The OpenFOAM boundary condition settings for velocity and pressure are given in Table 1. For all the boundary conditions supported by OpenFOAM, refer to OpenFOAM (2011).

**Table 1.** Boundary conditions for velocity and pressure

| Boundary | BC for velocity | BC for pressure |
|---|---|---|
| Inlet | fixedValue | zeroGradient |
| Outlet | zeroGradient | fixedValue (0) |
| Side walls | slip | zeroGradient |
| blade | movingWallVelocity | zeroGradient |

### 2.4 Turbulence model

The Spalart-Allmaras (SA) (Spalart and Allmaras, 1992) turbulence model was used to solve for the turbulent eddy-viscosity. In this model, the turbulent eddy-viscosity is given by:

$$\nu_t = \tilde{\nu} f_{v1}, \quad f_{v1} = \frac{X^3}{X^3 + C_{v1}^3}, \quad X \equiv \tilde{\nu}/\nu \quad (1)$$

Where $\nu$ is molecular viscosity and $\tilde{\nu}$ is a new variable given by the following equations,

$$\frac{D\tilde{\nu}}{Dt} = c_{b1}\tilde{S}\tilde{\nu} + \frac{1}{\sigma}\left[\nabla \cdot \left((\nu + \tilde{\nu})\nabla\tilde{\nu}\right) + c_{b2}(\nabla\tilde{\nu})^2\right] - c_{\omega 1}f_\omega\left[\frac{\tilde{\nu}}{d}\right]^2 \quad (2)$$

$$\tilde{S} \equiv S + \frac{\tilde{\nu}}{k^2 d^2} f_{v2}, \quad f_{v2} = 1 - \frac{X}{1 + Xf_{v2}} \quad (3)$$

In the SA turbulence model, $\tilde{\nu}$ is less expensive to compute than the coupled turbulent kinetic energy $k$ and dissipation rate $\epsilon$. The SA turbulence model was developed at Boeing and is often favored in aerodynamic applications (Javaherchi T 2010). The SA turbulence model requires boundary condition on the variable $\tilde{\nu}$. In our simulation the boundary condition settings of $\tilde{\nu}$ on different patches are given in the following table.

**Table 2.** Boundary conditions for $\tilde{v}$

| Patch | BC for $\tilde{v}$ |
|---|---|
| Inlet | FixedValue ($1.85\text{e-}4 \text{ m}^2/s$) |
| Outlet | zeroGradient |
| Side walls | zeroGradient |
| blade | FixedValue (0) |

## 3. RESULTS AND DISCUSSION

### 3.1 Pressure distributions

Comparisons of the NREL experimental data and the computed pressure distributions for 5m/s, 10m/s and 21m/s inlet wind speeds at three span sections, r/R = 30%, 47% and 80%, are shown in Figure 8, Figure 9 and Figure 10, respectively.

For the case with a 5m/s inlet wind speed, good agreement is achieved for all three span locations, as shown in Figure 8. This is due to the fact that at this low inlet wind speed the blade functions as designed and there is no boundary layer separation.

For the higher inlet wind speed of

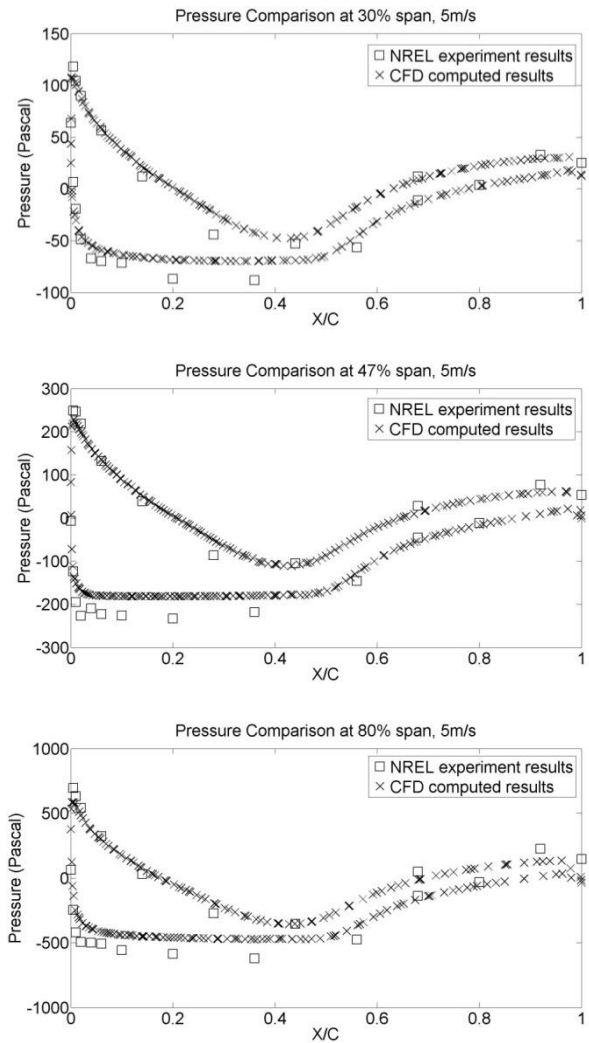

**Figure 8.** Comparison of the pressure distributions for the 5m/s case (a) at 30% span (b) at 47% span (c) 80% span.

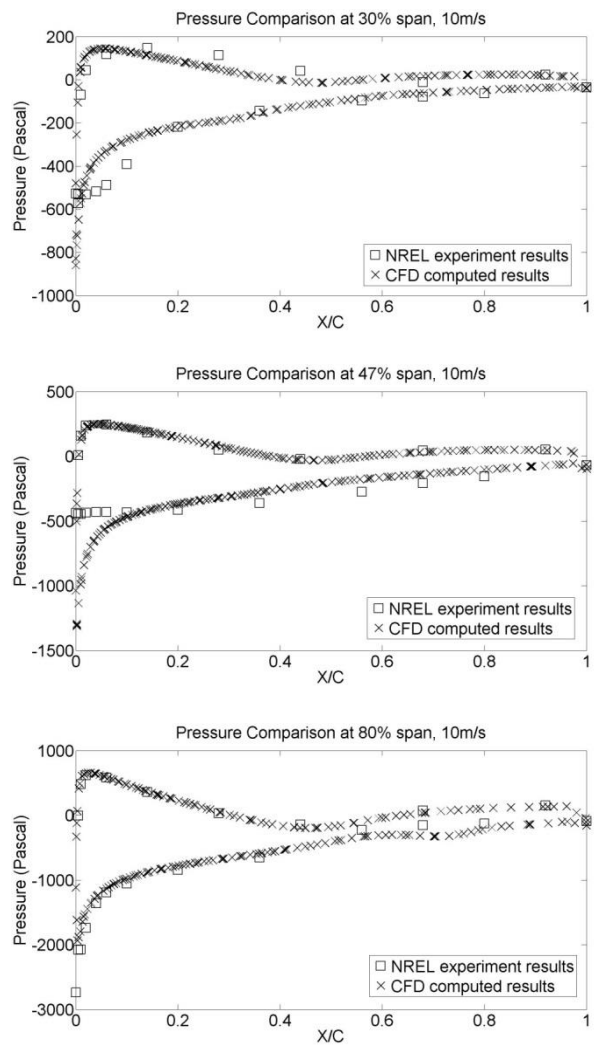

**Figure 9.** Comparison of the pressure distributions for the 10m/s case (a) at 30% span (b) at 47% span (c) 80% span.

10m/s, good agreement is also found at the 80% span location, as shown in Figure 9. However, the 30% and 47% span locations are predicted less well. The difficulty lies on the top surface of the airfoil (lower curve) at the leading edge. It will be shown later that these cross sections are experiencing separation. The CFD simulation overpredicts the pressure peak.

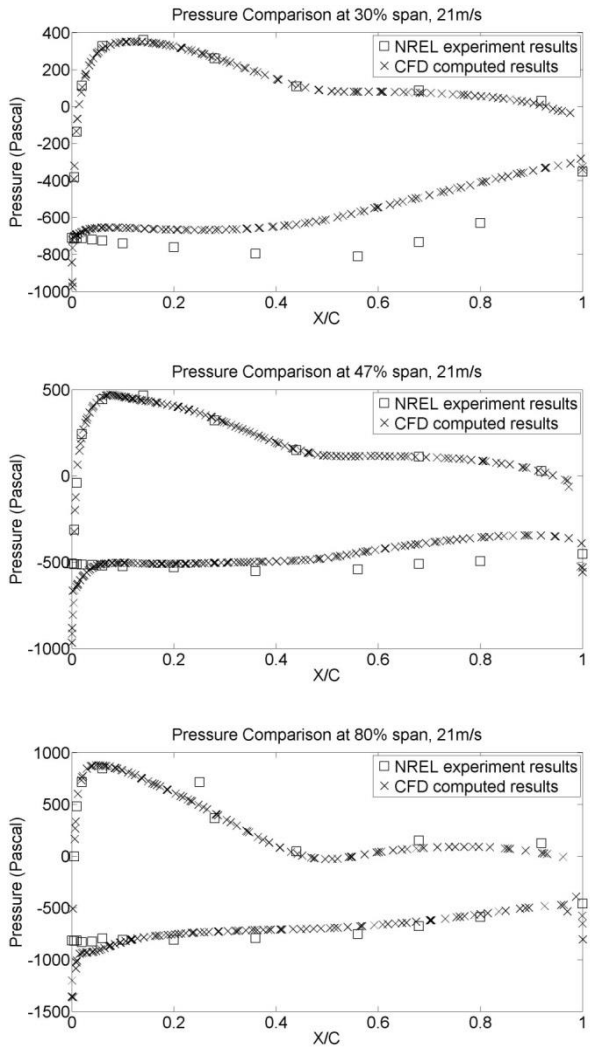

**Figure 10.** Comparison of the pressure distributions for the 21m/s case (a) at 30% span (b) at 47% span (c) 80% span.

For the highest inlet wind speed of 21m/s, pressure distributions are again over predicted at the leading edge for all three span locations, as shown in Figure 10. For this case the incoming wind speed is so large that the entire blade is under complete stall conditions with separation occurring at the leading edge of the blades.

### 3.2 Low speed shaft torque

In this section the low speed shaft torque (LSST) is computed for a series of simulations that all contain 10 million mesh cells but the inlet wind speed is varied from 5m/s to 21m/s. The results are shown in Figure 11, where it can be seen that the overall shape of the computed LSST curve is general agreement with the experimental LSST curve. After 10 m/s the blades are almost entirely stalled. The CFD predictions however predict a stronger stall, and less torque, than found in the experiments. This is likely a result of the turbulence model. There are no turbulence models which are known to predict this type of strong stall well.

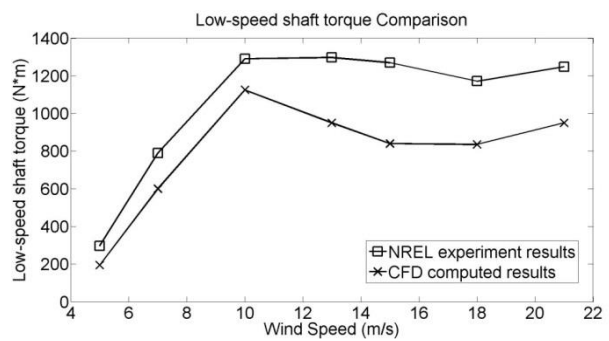

**Figure 11.** Comparison of low speed shaft torque for wind speeds of 5m/s, 7m/s, 10m/s, 13m/s, 15m/s, 18m/s and 21m/s.

Another way to analyze the stall effect is provided by Figure 12, which shows the limiting streamlines on the suction side of the blade for the inlet velocity of 5m/s, 10m/s and 21m/s. We can see that at 5m/s of inlet wind speed, the blade is operating as designed and has no stall effects. This is believed to be one explanation for the good agreement for the low inlet wind speed cases. At 10m/s, although the blade is stalling near the root, it is behaving fine near the tip where the blade has a lower angle of attack. The 21m/s case is completely stalled so it has a poor agreement.

**3.3 3D effects**

Streamlines and the second invariant of the velocity gradient tensor are used to examine the three-dimensionality of the flow on the rotating blades. Figure 13 shows the streamlines on the suction side of the blade for the 21 m/s case. All the streamlines originate at the root. Because of the strong stall, fluid is moving down the blade towards the tip. It is also moving towards the leading edge. When it reaches the leading edge it is swept off the blade in the separation shear layer.

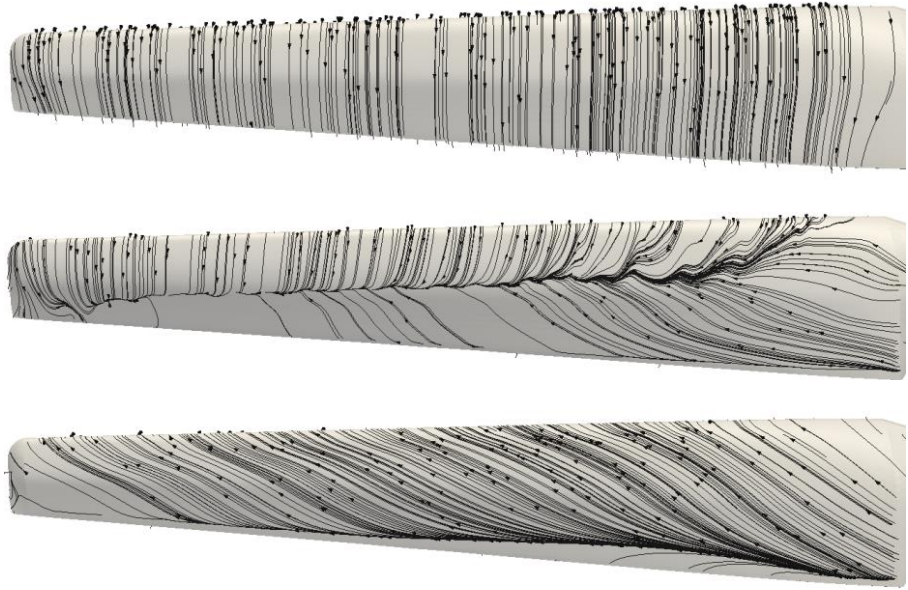

**Figure 12.** Limiting streamlines on the suction side of the blade for the inlet velocity of 5m/s, 10m/s and 21m/s

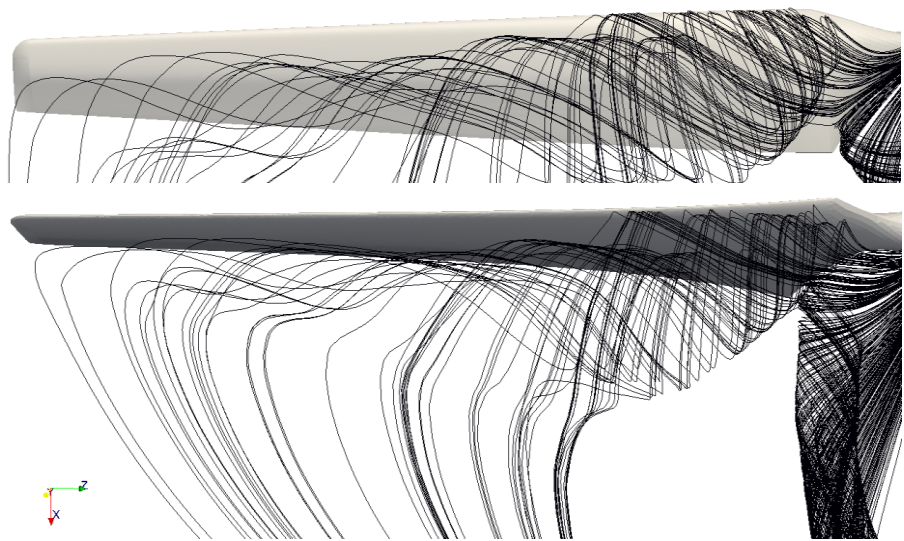

**Figure 13.** Two views of the streamlines on the suction side of the blade for 21 m/s case.

Figure 14 shows an iso-surface of the second invariant velocity gradient tensor at the value $Q = 0.3$ $1/s^2$ for the 10 m/s case. This invariant is a good indicator of vortices. In this case it clearly identifies the trailing tip vortices, and also an inner pair of trailing vortices that emanate from where the blades begin at the root. Figure 15 shows an iso-surface of the X velocity at $U_X = 8.1$m/s also for the 10 m/s inflow case.

## 4. CONCLUSIONS

A series of CFD simulations of the NREL Phase VI wind turbine with $0^o$ yaw angle and $3^o$ tip pitch angle at a rotation rate of 72 rpm were performed. Significant effort was made to refine the mesh at the leading edge and normal to the blade surface to accurately resolve the thin physical flow features in the velocity and pressure. Generally good agreement with the NREL experimental results was found for inlet wind speeds lower than 10m/s where the blades are not totally stalled. For inlet wind speeds higher than 10m/s, larger differences are observed between the simulations and experiments. It is likely these differences are due to the limitations of the turbulence model.

The importance of good mesh quality to a successful and accurate CFD prediction was analyzed by testing the CFD cod performance with three different mesh resolutions. The low-speed shaft torque comparison shows that the computed CFD results are able to capture the basic trends of the NREL experimental results even though some quantitative differences are observed. The three-dimensionality of the flow under separation conditions is shown to be very significant.

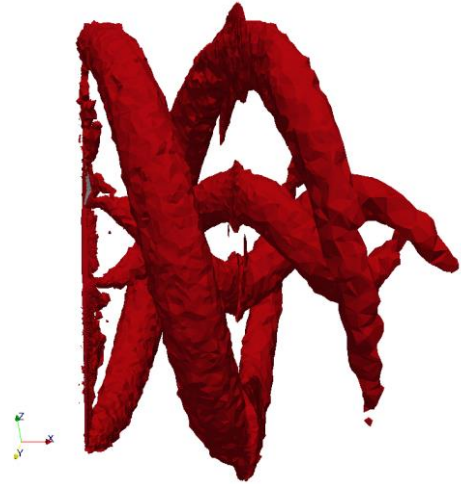

**Figure 14.** Iso-surface of the second invariant velocity gradient tensor at $Q = 0.3$ $1/s^2$, for the 10m/s case. This identifies the trailing vortices at the tip and the root.

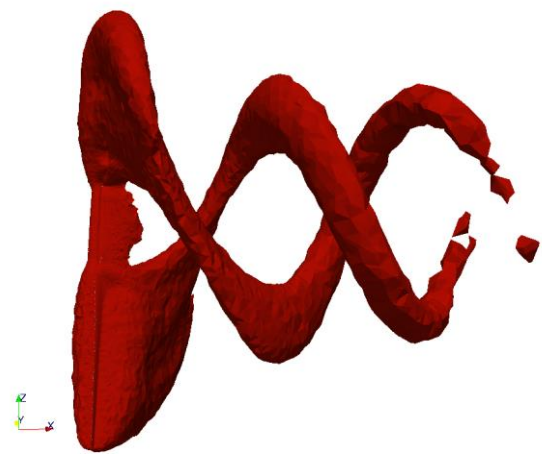

**Figure 15.** Iso-surface of X velocity at $U_X = 8.1$ m/s, for the 10m/s case.